# Parallel algorithms for SAT in application to inversion problems of some discrete functions

Alexander Semenov, Oleg Zaikin, Dmitry Bespalov, Mikhail Posypkin

**Abstract**—In this article we consider the inversion problem for polynomially computable discrete functions. These functions describe behavior of many discrete systems and are used in model checking, hardware verification, cryptanalysis, computer biology and other domains. Quite often it is necessary to invert these functions, i.e. to find an unknown preimage if an image and algorithm of function computation are given. In general case this problem is computationally intractable. However, many of it's special cases are very important in practical applications. Thus development of algorithms that are applicable to these special cases is of importance. The practical applicability of such algorithms can be validated by their ability to solve the problems that are considered to be computationally hard (for example cryptanalysis problems). In this article we propose the technology of solving the inversion problem for polynomially computable discrete functions. This technology was implemented in distributed computing environments (parallel clusters and Grid-systems). It is based on reducing the inversion problem for the considered function to some SAT problem. We describe a general approach to coarse-grained parallelization for obtained SAT problems. Efficiency of each parallelization scheme is determined by the means of a special predictive function. The proposed technology was validated by successful solving of cryptanalysis problems for some keystream generators. The main practical result of this work is a complete cryptanalysis of keystream generator A5/1 which was performed in a Grid system specially built for this task.

## 1 INTRODUCTION

Let $\{0,1\}^n, n \in \mathrm{N}_1$, be the set of all possible binary sequences of the length $n$. We also use the notation $\{0,1\}^* = \bigcup_{n \in \mathrm{N}_1} \{0,1\}^n$. This work focuses on the families of discrete functions of the form

$$f = \{f_n\}_{n \in \mathrm{N}_1}, \; f_n : \{0,1\}^n \to \{0,1\}^*.$$

- Alexander Semenov, Oleg Zaikin, Dmitry Bespalov are with the Laboratory of Discrete Analysis and Applied Logic, Institute for System Dynamics and Control Theory of Siberian Branch of Russian Academy of Sciences, Lermontov str. 134, Irkutsk, 664033, Russia.
  E-mail: biclop@rambler.ru, oleg.zaikin@icc.ru, bespalov@altrixsoft.com
- Mikhail Posypkin is with the Centre for Grid Technologies and Distributed Computing, Institute for Systems Analysis of Russian Academy of Sciences, pr. 60-letiya Oktyabrya 9, Moscow, 117312, Russia
  E-mail: posypkin@isa.ru

We consider the class formed by the families of discrete functions for which the following conditions are satisfied:

1) for every $n \in \mathrm{N}_1$ the function $f_n$ is defined everywhere on $\{0,1\}^n$ (we denote this fact as $dom\, f_n = \{0,1\}^n$);
2) there exists a program $M_f$ for deterministic Turing machine which computes an arbitrary function of the family $f$;
3) the time complexity of the program $M_f$ increases with the increase of $n$ as a polynomial in $n$.

Hereafter we write $f_n \in f$ to indicate the fact that the discrete function $f_n$ belongs to a family with the properties 1–3. For a discrete function $f_n \in f$ given in the form of a pair $(M_f, n)$ and a word $y \in range\, f_n$, the problem we are interested in is to find $x \in dom\, f_n$, such that $y = f_n(x)$. We call this problem the inversion problem for the function $f_n$ in the point $y$. The general inversion problem for family $f$ with properties 1–3 we denote by $\mathrm{Inv}(f)$.

In the present article we describe an approach that is based on the possibility to reduce effectively the problem $\mathrm{Inv}(f)$ to SAT.

In the process of reduction of the problem $\mathrm{Inv}(f)$ to a SAT problem there is a possibility to single out from the set of variables of the obtained CNF the subset corresponding to the "input variables" of the considered function. This is fundamental for constructing of decompositions of SAT problems into SAT problems of lower dimension with their subsequent solving in distributed computing environments. We show that the use of this simple principle gives good results for the inversion problems of some discrete functions used in cryptography.

Let us give a brief outline of the article. In the second section we give basic notions of the theory of discrete functions and briefly describe a technology for reducing inversion problems of the functions computable in polynomial time to SAT problems, focusing on the functions used in cryptography.

In the third section, technology of coarse-grained parallelism that we use for solving SAT problems, is described. Using this technology we decompose given SAT problem into a family of SAT problems of lower dimension. Such decompositions can be performed in different ways. We are interested in selecting a decomposition that is good in terms of overall computing time. This is performed by solving optimization problem for a special predictive function.

In the fourth section, we describe some modifications intended to improve efficiency of a basic SAT solver for the problem $\mathrm{Inv}(f)$. Here we also give the results of a successful cryptanalysis of some keystream generators performed on a low-performance computing cluster.

The fifth section is entirely devoted to the use of the technology presented in this paper for solving the problem of cryptanalysis of the keystream generator A5/1 using a Grid system specially constructed for this purpose.

## 2 REDUCING THE DISCRETE FUNCTIONS INVERSION PROBLEMS TO SAT PROBLEMS

### 2.1 Basic notions

Hereafter $X = \{x_1, \ldots, x_n\}$ denotes the set of Boolean variables. Let $L(x_1, \ldots, x_n)$ be an arbitrary propositional formula. We denote by $L(\alpha_1, \ldots, \alpha_n) = \beta$ the fact that the result of substitution

$$x_1 = \alpha_1, \ldots, x_n = \alpha_n,$$

$\alpha_i \in \{0,1\}$, $i \in \{1, \ldots, n\}$, into the formula $L(x_1, \ldots, x_n)$ is $\beta \in \{0,1\}$.

The expressions of the form $L(x_1, \ldots, x_n) = 0$, $L(x_1, \ldots, x_n) = 1$ are called Boolean equations (see [1]). For a fixed $\beta \in \{0,1\}$ a solution of the equation $L(x_1, \ldots, x_n) = \beta$ is a vector $(\alpha_1, \ldots, \alpha_n) \in \{0,1\}^n$, such that $L(\alpha_1, \ldots, \alpha_n) = \beta$. If such a vector does not exist we say that the Boolean equation does not have a solutions.

The terms $x_i, \overline{x_i}, i \in \{1, \ldots, n\}$, are called literals over $X$. The literals $x$ and $\overline{x}$ are called complementary literals. A clause over $X$ is an arbitrary disjunction of literals over $X$, which does not have repetitive and complementary literals. Conjunctive normal form (CNF) over $X$ is an arbitrary conjunction of different clauses over $X$.

Let $C(x_1, \ldots, x_n)$ (shortly $C$) be an arbitrary CNF over the set of Boolean variables $X = \{x_1, ..., x_n\}$. The vector $(\alpha_1, ..., \alpha_n) \in \{0,1\}^n$ is called a satisfying assignment of $C$, if $C(\alpha_1, ..., \alpha_n) = 1$. A CNF for which there exists a satisfying assignment is called a satisfiable CNF, otherwise it is called unsatisfiable. The problem of deciding satisfiability of an arbitrary CNF as well as the problem of search of a satisfying assignment for an arbitrary satisfiable CNF are the problems we consider below.

S.A. Cook showed in [2], that the process of executing a program $M$, which stops on an arbitrary input, on a Turing machine with the input alphabet $\Sigma = \{0,1\}$ can be represented by a system of Boolean equations.

Let $f$ be a family of discrete functions from the class defined above. An arbitrary function $f_n \in f$ given by a pair $(M_f, n)$ will be considered as a function of Boolean variables from the set $X = \{x_1, \ldots, x_n\}$. The set $X$ we call the set of input variables of the function $f_n$. According to [2] there exists an algorithm with a time complexity bounded by polynomial in $n$ which given a pair $(M_f, n)$, transforms the problem of inversion of $f_n$ in an arbitrary point $y \in range\, f_n$ into the problem of finding solutions of the equation of the type

$$C(x_1, \ldots, x_{q(n)}) = 1. \quad (1)$$

Here $q(\cdot)$ is some polynomial, and $C(x_1, \ldots, x_{q(n)})$ is a satisfiable CNF over the set of Boolean variables $\{x_1, \ldots, x_{q(n)}\}$. Further we will write "CNF encoding discrete function inversion problem" meaning CNF $C(x_1, \ldots, x_{q(n)})$ and "equation encoding discrete function inversion problem" meaning equation $C(x_1, \ldots, x_{q(n)}) = 1$.

It should be particularly noted that the procedure for reducing the inversion problem to the search for solutions of Boolean equations must be parsimonious (see [3], [4]). That is, the number of solutions of (1) must coincide with the number of preimages of $y \in range\, f_n$, and a procedure of an effective transition from an arbitrary solution of (1) to the corresponding preimage must exist. This is essential for an effective inversion of discrete functions in practice.

Not all procedures of propositional encoding are parsimonious. However, it is not difficult to show that well-known Tseitin transformations have this property. These transformations were proposed by G.S. Tseitin in 1968 in [5] (reprinted in [6]). Next, we describe the use of Tseitin transformations for the problem of parsimonious reducing of Boolean equations to normal forms.

These transformations were described (explicitly or implicitly) in a number of sources (e.g., [7], [8]) where an original function is usually represented by a Boolean circuit $S(f_n)$ over an arbitrary complete basis, for example $\{\&, \neg\}$. Each variable from $X$ corresponds to one of $n$ inputs of $S(f_n)$. For each logic gate $G$ some new auxiliary variable $v(G)$ is introduced. We denote the set of all auxiliary variables as $V$. Every AND-gate $G$ is encoded by CNF-representation of Boolean function $v(G) \leftrightarrow u \,\&\, w$. Every NOT-gate $G$ is encoded by CNF-representation of Boolean function $v(G) \leftrightarrow \neg u$. Here $u$ and $w$ are variables corresponding to inputs of $G$. CNF encoding $S(f_n)$ is

$$\left( \underset{G \in S(f_n)}{\&} C(G) \right) \cdot y_1 \cdot \ldots \cdot y_m,$$

where $C(G)$ is CNF encoding gate $G$ and $y_1, \ldots, y_m$ are variables corresponding to outputs of $S(f_n)$.

In our opinion for the problems considered in this paper it is more convenient to construct Boolean equations that encode considered algorithms directly and not to use Boolean circuits for intermediate representation of these algorithms (see Section 2.2).

We consider the problem of finding solutions of Boolean equations in the following general formulation.

$$F\left( h_1\left( x_1^1, \ldots, x_{r_1}^1 \right), \ldots, h_s\left( x_1^s, \ldots, x_{r_s}^s \right) \right) = 1. \quad (2)$$

The propositional formulae $h_i\left( x_1^i, \ldots, x_{r_i}^i \right), i \in \{1, \ldots, s\}$, define some (composite in general case) Boolean functions. Let

$$X = \{x_1, \ldots, x_n\} = \bigcup_{i=1}^{s} \left\{ x_1^i, \ldots, x_{r_i}^i \right\}.$$

Consider the Boolean equation

$$C\left( u_1 \leftrightarrow h_1 \right) \cdot F_{h_1 \to u_1}(x_1, \ldots, x_n, u_1) = 1. \quad (3)$$

Here by $C(u_1 \leftrightarrow h_1)$ we denote a CNF-representation of the Boolean function

$$u_1 \leftrightarrow h_1\left( x_1^1, \ldots, x_{r_1}^1 \right)$$

over $\left\{ x_1^1, \ldots, x_{r_1}^1, u_1 \right\}$, and by

$$F_{h_1 \to u_1}(x_1, \ldots, x_n, u_1)$$

we denote the propositional formula obtained by replacing one or several (perhaps all) formulae $h_1\left( x_1^1, \ldots, x_{i_1}^1 \right)$ in (2) by the literal $u_1$.

The transition from the equation (2) to the equation (3) is one iteration of the Tseitin transformations in application to the Boolean equations.

Let $\Phi_1$ and $\Phi_2$ are the sets of solutions of the equations of (2) and (3) respectively. It is not difficult to show that the reduction from (2) to (3) described above is parsimonious because $\Phi_1$ and $\Phi_2$ are either simultaneously empty or there exists one-to-one correspondence between them. Implicitly this fact is mentioned in [8].

## 2.2 Logical cryptanalysis. Propositional encoding of the keystream generator A5/1

The concept of reducing the problems of cryptanalysis to the problems of finding solutions of Boolean equations (in the form of SAT problems) was first formulated in [9]. One of the first practical implementations of this idea was given in [10]. In that paper the problem of cryptanalysis of the DES cipher was formulated as a SAT problem.

Next, we consider the keystream generator A5/1 used to encrypt traffic in GSM networks. A lot of attacks on the cipher A5/1 are described, however it is still actively used. A possible reason of using A5/1 is the lack of convincing experimental results of its cryptanalysis. We consider the problem of cryptanalysis of the generator A5/1 on the basis of a known keystream. The problem is to find the secret key using some fragment of the keystream and a known algorithm of its generation (see [11]). The description of the generator A5/1 (see Fig. 1) was taken from the paper [12]. According to [12] generator A5/1 contains three linear feedback shift register (LFSR, see, e.g., [11]), given by the following connection polynomials: LFSR 1: $X^{19}+X^{18}+X^{17}+X^{14}+1$; LFSR 2: $X^{22} + X^{21} + 1$; LFSR 3: $X^{23} + X^{22} + X^{21} + X^8 + 1$.

The secret key of A5/1 generator is the initial contents of LFSRs 1–3 (64 bits). In each unit of time $\tau \in \{1, 2, \ldots\}$ ($\tau = 0$ is reserved for the initial state) two or three registers are shifted. The register with number $r$, $r \in \{1, 2, 3\}$, is shifted if $\chi_r^\tau (b_1^\tau, b_2^\tau, b_3^\tau) = 1$, and is not shifted if $\chi_r^\tau (b_1^\tau, b_2^\tau, b_3^\tau) = 0$. By $b_1^\tau$, $b_2^\tau$, $b_3^\tau$ we denote here the values of the clocking bits at the current unit of time. The clocking bits are 9-th, 30-th and 52-nd. Corresponding cells in Fig. 1 are black. The function $\chi_r^\tau (\cdot)$ is defined as follows

$$\chi_r^\tau (b_1^\tau, b_2^\tau, b_3^\tau) = \begin{cases} 1, & b_r^\tau = \text{majority}\,(b_1^\tau, b_2^\tau, b_3^\tau) \\ 0, & b_r^\tau \neq \text{majority}\,(b_1^\tau, b_2^\tau, b_3^\tau) \end{cases}$$

where $\text{majority}\,(A, B, C) = A \cdot B \vee A \cdot C \vee B \cdot C$.

In each unit of time the values in the leftmost cells of the registers are added mod 2, the resulting bit is the bit of the keystream.

Thus, we can see that the generator A5/1 updates the content of each of the registers' cells as a result of conditional shifts: if the shift does not occur, then a new configuration of a register does not differ from the old one, otherwise values of all cells of the register are updated. Hence with each cell at each unit of time we can associate a Boolean equation linking a new state of the cell with the previous one. Let variables $x_1, \ldots, x_{64}$ encode the secret key of generator A5/1 ($x_i$ corresponds to cell with number $i \in \{1, \ldots, 64\}$). By $x_1^1, \ldots, x_{64}^1$ we denote variables encoding cells' state in the moment of time $\tau = 1$. System of equations which links these two sets of variables is:

$$\begin{cases} \left(x_1^1 \leftrightarrow x_1 \cdot \overline{\chi_1^1} \vee (\oplus_{i \in I} x_i) \cdot \chi_1^1\right) = 1 \\ \left(x_2^1 \leftrightarrow x_2 \cdot \overline{\chi_1^1} \vee x_1 \cdot \chi_1^1\right) = 1 \\ \ldots\ldots\ldots\ldots\ldots\ldots\ldots\ldots \\ \left(x_{20}^1 \leftrightarrow x_{20} \cdot \overline{\chi_2^1} \vee (\oplus_{j \in J} x_j) \cdot \chi_2^1\right) = 1 \\ \left(x_{21}^1 \leftrightarrow x_{21} \cdot \overline{\chi_2^1} \vee x_{20} \cdot \chi_2^1\right) = 1 \\ \ldots\ldots\ldots\ldots\ldots\ldots\ldots\ldots \\ \left(x_{42}^1 \leftrightarrow x_{42} \cdot \overline{\chi_3^1} \vee (\oplus_{k \in K} x_k) \cdot \chi_3^1\right) = 1 \\ \left(x_{43}^1 \leftrightarrow x_{43} \cdot \overline{\chi_3^1} \vee x_{42} \cdot \chi_3^1\right) = 1 \\ \ldots\ldots\ldots\ldots\ldots\ldots\ldots\ldots \\ \left(x_{64}^1 \leftrightarrow x_{64} \cdot \overline{\chi_3^1} \vee x_{63} \cdot \chi_3^1\right) = 1 \\ \left(g^1 \leftrightarrow x_{19}^1 \oplus x_{41}^1 \oplus x_{64}^1\right) = 1 \end{cases} \quad (4)$$

where $I = \{14, 17, 18, 19\}$, $J = \{40, 41\}$, $K = \{49, 62, 63, 64\}$ and $g^1$ is the first bit of keystream.

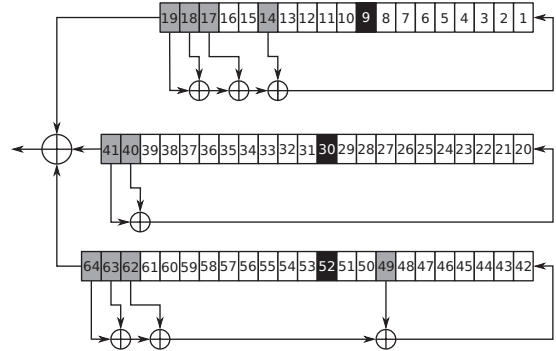

Fig. 1. Scheme of the generator A5/1.

Let $g^1, \ldots, g^L$ be the first $L$ bits of the keystream of A5/1. To the each bit $g^i, i \in \{1, \ldots, L\}$ we associate a system of the form (4). To find the secret key it is sufficient to find a common solution of these systems. The prob-lem of finding of this common solution can be reduced by the means of Tseitin transformations to the problem of finding a satisfying assignment of a satisfiable CNF.

## 3 COARSE-GRAINED PARALLELIZATION OF SAT PROBLEMS ENCODING DISCRETE FUNCTIONS INVERSION PROBLEMS

In this section we describe a technology for solving SAT problems in distributed computing systems (hereinafter DCS). Such systems consist of sets of computing nodes connected by a communication network. Each node of a DCS has one or several processors. Typical examples of DCS are computing clusters which have become widespread in recent years. The elementary computational units of modern DCS are cores of processors.

One of the first works on parallel algorithms for solving SAT problems is the article [13]. This work describes a technique of parallelization of the Davis-Putnam procedure (see [14] with interprocessor data exchange aimed to achieve uniform loading of the processors. Similar ideas are basic for those modern SAT solvers which use interprocessor exchange of conflict clauses (see, e.g., [15]).

In the present work a different approach to parallelization of algorithms for solving SAT problems is proposed. This approach is primarily oriented to solving the problems from $\mathrm{Inv}\,(f)$. We show the principal possibility of using preliminary calculations to determine good parameters of a decomposition of the search domain into disjoint subdomains. After decomposition the obtained subdomains are processed by isolated processors. In this sense the main results of this article belong entirely to the field of the coarse-grained parallelism (see, e.g., [16]).

We consider an arbitrary CNF $C$ over the set of Boolean variables $X = \{x_1, \ldots, x_n\}$ and select in the set $X$ some subset

$$X' = \{x_{i_1}, \ldots, x_{i_d}\}, \{i_1, \ldots, i_d\} \subseteq \{1, \ldots, n\},$$

where $d \in \{1, \ldots, n\}$. We call $X' = \{x_{i_1}, \ldots, x_{i_d}\}$ a decomposition set and $d$ is the power of the decomposition set. To the decomposition set $X'$, $|X'| = d$, we associate the set $Y(X') = \{Y_1, \ldots, Y_K\}$ consisting from $K = 2^d$ different binary vectors of the length $d$, each of which is a vector of values of the variables $x_{i_1}, \ldots, x_{i_d}$. By $C_j = C|_{Y_j}$, $j = 1, \ldots, K$, we denote the CNF obtained after substitutions of the values from the vectors $Y_j$ to $C$. A decomposition family generated from the CNF $C$ by the set $X'$, is the set $\Delta_C(X')$, formed by the following CNFs:

$$\Delta_C(X') = \{C_1 = C|_{Y_1}, \ldots, C_K = C|_{Y_K}\}.$$

It is not difficult to see that any truth assignment $\alpha \in \{0,1\}^n$ satisfying $C$ ($C|_\alpha = 1$) coincides with some vector $Y^\alpha \in Y(X')$ in the components from $X'$ and coincides with some satisfying assignment of the CNF $C|_{Y^\alpha} \in \Delta_C(X')$ in the remaining components. In this case the CNF $C$ is unsatisfiable if and only if all the CNF in $\Delta_C(X')$ are unsatisfiable. Therefore, the SAT problem for the original CNF $C$ is reduced to $K$ SAT problems for CNFs from the set $\Delta_C(X')$. For processing the set $\Delta_C(X')$ as a parallel task list a DCS can be used.

Note that the idea of such parallelization itself is not new. A similar in spirit approach to SAT problems is presented in [17]. The main novelty of our approach consists in the described further technique of search of decomposition sets with "good" properties. This technique is based on a simple procedure of statistical prediction.

In the case of arbitrary SAT problems it is not clear how to form decomposition sets. However, below we will show that for the SAT problem encoding the problem of inversion of a discrete function $f_n$ good candidates for the role of decomposition sets are subsets of the set of input variables of $f_n$.

If we take as a decomposition set the whole set of input variables of $f_n$, then every SAT problem in the obtained family is simple, but the number of these problems is usually very big. Thus there arises the following problem of improvement of a decomposition set. Let $X'$ be some decomposition set (e.g., the set of input variables of $f_n$). We need to construct a set $\tilde{X} \subset X'$, for which there exist some conclusions about smaller total processing time of the list $\Delta_C(\tilde{X})$.

If the power of the set $\Delta_C(\tilde{X})$ is too high, then the prediction of the time of parallel processing of $\Delta_C\left(\tilde{X}\right)$ can be computed based on the average solving time of SAT problems for some CNFs that are randomly chosen (with uniform distribution) from $\Delta_C\left(\tilde{X}\right)$. By $Q\left(\tilde{X}\right)$ we denote the size of this sample.

Introduce a parameter $R$ to distinguish two situations: whether it is necessary to form a random sample or not.

To each set $\tilde{X}$, $\tilde{X} \subseteq X'$, such that $2^{|\tilde{X}|} > R$, we associate the set of vectors $\left\{Y'_1, \ldots, Y'_{Q(\tilde{X})}\right\}$, selected from $Y\left(\tilde{X}\right)$ with uniform distribution and the set of CNFs

$$\theta_C\left(\tilde{X}\right) = \left\{C'_1 = C|_{Y'_1}, \ldots, C'_{Q(\tilde{X})} = C|_{Y_{Q(\tilde{X})}}\right\}.$$

To each $\tilde{X}$, $\tilde{X} \subseteq X'$, such that $2^{|\tilde{X}|} \leq R$, we associate the set $Y\left(\tilde{X}\right)$ and the set of CNFs $\theta_C\left(\tilde{X}\right) = \Delta_C\left(\tilde{X}\right)$.

Thus, to an arbitrary set $\Omega$, $\Omega \subseteq 2^{X'}$, of the choice alternatives of $\tilde{X}$ from $X'$, the set

$$\Theta_C\left(\Omega\right) = \left\{\theta_C\left(\tilde{X}\right)\right\}_{\tilde{X} \in \Omega}$$

is put in correspondence.

Hereinafter we consider SAT solvers based on DPLL (see [18]). Using properties of this algorithm we construct the procedure for predicting time of parallel solving of SAT problems which encode discrete functions inversion problems.

Let $S$ be some SAT solver. Denote by $t\left(C'\right)$ the time of work of the SAT solver $S$ on an arbitrary CNF $C'$. Consider the function $\tau_S : \Theta_C\left(\Omega\right) \to N_1$,

$$\tau_S\left(\theta_C\left(\tilde{X}\right)\right) = \sum_{C' \in \theta_C(\tilde{X})} t\left(C'\right).$$

However, for some $\tilde{X}$ (e.g., if $\tilde{X}$ consists of one variable) CNF from $\theta_C\left(\tilde{X}\right)$ can be very difficult for the SAT solver. In this case the time required for computing the corresponding value of the predictive function may exceed reasonable limits. To take this fact into account we introduce a special function $g\left(C\right)$.

Suppose that in accordance with the rules above the set $\Theta_C\left(\Omega\right)$ is constructed. We define the predictive function as follows.

$$T\left(\theta_C\left(\tilde{X}\right)\right) = \begin{cases} \frac{2^{|\tilde{X}|}}{Q(\tilde{X})} \cdot \tau_S\left(\theta_C\left(\tilde{X}\right)\right), \\ \text{if } 2^{|\tilde{X}|} > R, \tau_S\left(\theta_C\left(\tilde{X}\right)\right) < g(C) \\ \\ \tau_S\left(\theta_C\left(\tilde{X}\right)\right), \\ \text{if } 2^{|\tilde{X}|} \leq R, \tau_S\left(\theta_C\left(\tilde{X}\right)\right) < g(C) \\ \\ \infty, \text{ if } \tau_S\left(\theta_C\left(\tilde{X}\right)\right) \geq g(C) \end{cases}$$

Notation $T\left(\theta_C\left(\tilde{X}\right)\right) = \infty$ means that the function is not defined in $\theta_C\left(\tilde{X}\right)$. The value $T\left(\theta_C\left(\tilde{X}\right)\right)$ is the prediction of the time required for a sequential processing of the list $\Delta_C\left(\tilde{X}\right)$. Thus the problem of constructing a "good" decomposition set is reduced to the problem of minimizing the function $T\left(\cdot\right)$ on the set $\Theta_C\left(\Omega\right)$. Knowing the global minimum of $T\left(\cdot\right)$ on $\Theta_C\left(\Omega\right)$ we can make a conclusion about the possibility of parallel solving of a SAT problem for the CNF $C$ in "a reasonable time".

**Theorem 1** *Consider the problem* $\text{Inv}\left(f\right)$. *Let $C$ be a CNF encoding the problem of inversion of function $f_n, f_n \in f$ in an arbitrary point $y \in range f_n$. Then there exist sets $\Omega$, $\Theta_C\left(\Omega\right)$ : $|\Theta_C\left(\Omega\right)| \leq 2^n$ and function $T\left(\cdot\right)$ such that the domain of $T\left(\cdot\right)$ is nonempty and global minimum of $T\left(\cdot\right)$ on $\Theta_C\left(\Omega\right)$ can be found in time*

$$O\left(|C| \cdot |\Theta_C\left(\Omega\right)|\right).$$

*Proof:* Suppose that $\Omega$ contains the set $X$ of input variables of function $f_n$. Here we assume that the CNF $C$ encoding a corresponding inversion problem is constructed in accordance with the principles described in Section 2.1. It is known (see, e.g., [19]) that DPLL remains complete for $C$ even if the set of decision variables (see [20]) is limited to $X$. Thus assigning values to all the variables from X will result in inference of either a satisfying assignment of $C$ or a contradiction (conflict) through the unit propagation (UP, see [21]). However generally the complexity of this process has an upper bound of the kind $O\left(|C|\right)$. Therefore it is always possible to construct some function $g\left(\cdot\right)$, $g\left(C\right) = O\left(|C|\right)$ such that for an arbitrary sample $\theta_C(X)$ the value of function $T(\theta_C(X))$ will be defined.

Next, we describe an algorithm based on the principle of "dynamic programming" for solving the problem of minimizing the function $T(\cdot)$ on the set $\Theta_C(\Omega)$.

Let the initial decomposition set $X'$ be the set of the input variables of $f_n$ and $2^{|X'|} > R$. We construct the set $\theta_C(X')$ and find the value $T(\theta_C(X'))$. As already mentioned, this value can be found effectively. Consequently, the function $T(\cdot)$ is a defined at least in $\theta_C(X')$. Next, we will try to sequentially improve the value of the function $T(\cdot)$ on some fixed set $\Omega \subset 2^{X'}$.

By $i$ we denote the iteration number of the algorithm. The results of the initial ($i = 0$) iteration step are the values $\tau_S(\theta_C(X'))$ and $T(\theta_C(X'))$. At each iteration step $i \geq 1$ we compute the values of the functions $\tau_S\left(\theta_C\left(\tilde{X}\right)\right)$ and $T\left(\theta_C\left(\tilde{X}\right)\right)$ for the corresponding set $\tilde{X} \in \Omega$. We denote them as $\tau^i$ and $T^i$ respectively. For each $i \geq 1$ in the process of calculations we perform frequent checks to determine whether the current value of $T^i$ has exceeded the value $T^{i-1}$ which was found at previous iteration step. If this takes place then further calculations are useless since the calculated value $T^i$ won't be better than $T^{i-1}$. Then we interrupt the calculations and go to the next iteration step. For example we denote by $\tau^0$ the value $\tau_S(\theta_C(X'))$ and calculate $T\left(\theta_C\left(\tilde{X}\right)\right)$ for some $\tilde{X}, \tilde{X} \subset X'$. If the inequality

$$\tau_S\left(\theta_C\left(\tilde{X}\right)\right) > 2^{|X'|-|\tilde{X}|} \cdot \tau^0$$

holds and $2^{|\tilde{X}|} > R$, $Q(\tilde{X}) = Q(X')$ then $T\left(\theta_C\left(\tilde{X}\right)\right) > T(\theta_C(X'))$. Obviously, in this case the global minimum of $T(\cdot)$ on $\Theta_C(\Omega)$ cannot be achieved in $\theta_C\left(\tilde{X}\right)$. If in the process of calculating $\tau_S\left(\theta_C\left(\tilde{X}\right)\right)$ the bound $g(C)$ is exceeded, then the value of $T(\cdot)$ in $\theta_C\left(\tilde{X}\right)$ is not defined.

The result of the described procedure of iterative improvement of the value of the function $T(\cdot)$ on $\Theta_C(\Omega)$ is a set $X_* \in 2^{X'}$, such that the value of $T(\theta_C(X_*))$ is minimal (among all $\tilde{X} \in \Omega$). The number of necessary iterations is $|\Theta_C(\Omega)|$ and upper bound for the time of each iteration is $O(|C|)$. □

Note that even if $X'$ is the set of input variables of a considered function, processing whole set $2^{X'}$ for practically important problems is unfeasible. Therefore, the peculiarities of the original formulation should be taken into account and various heuristics should be used to form $\Omega$ in each particular case. Of course, there is no guarantee that a decomposition set better (in terms of total computing time) then $X_*$ doesn't exist. However the examples described in Section 4 show practical effectiveness of the proposed method of predictive functions even with quite simple techniques of constructing of sets $\Omega$.

Further, we assume that we have some decomposition set $X_*$ with a good value $T(\theta_C(X_*))$ for some CNF $C$. The decomposition family generated by the set $X_*$ we denote by $\Delta_*(C) = \{C_1, \ldots, C_K\}$, $K = 2^{|X_*|}$. Suppose that the considered DCS has $M$ computing cores. The following two cases are possible.

1) $K \leq M$, i.e. the number of CNF in the family $\Delta_*(C)$ does not exceed the number of cores of the DCS. In this case for any CNF from the family $\Delta_*(C)$ the SAT problem is solved on a separate core. In practice this situation is very rare.
2) $K > M$—the number of CNF in the family $\Delta_*(C)$ is greater then the number of cores in the DCS.

The situation described in 2 is the most typical for inversion problems of the cryptographic functions. In this case the decomposition family $\Delta_*(C)$ is considered as a task list that is processed in parallel according to the following scheme. Let us put the CNFs of the family $\Delta_*(C)$ in some order. We call an arbitrary CNF from $\Delta_*(C)$ locked if at the current moment of time the SAT problem for it has either been solved or is being solved on some core of the DCS. The other CNFs are called free. We select first $M$ CNFs $C_1, \ldots, C_M$ from the family $\Delta_*(C)$. For each of the selected CNFs we solve the SAT problem on a separate core of the DCS. Once some core is released we launch the procedure of solving of the SAT problem for the first free CNF of the family $\Delta_*(C)$ on this core. This process continues un-til a satisfying assignment for some CNF from $\Delta_*(C)$ is found, or until the unsatisfiability of all CNFs from $\Delta_*(C)$ is proven.

In the next section we describe a somewhat different strategy of processing the list $\Delta_*(C)$, which significantly reduces the cost of the transfer of tasks over a network.

## 4 CRYPTANALYSIS OF SOME KEYSTREAM GENERATORS ON A COMPUTING CLUSTER

### 4.1 Adjustments of a SAT solver to solving the problems of cryptanalysis in distributed computing environments

In all experiments described further we use a modified SAT solver Minisat-C v1.14.1 (see [22]). The first stage of modification consists in changing the decision variable selection procedure (see [20]) implemented in Minisat. Namely, a procedure of assignment of initial activity (different from zero) for those variables in the CNF which correspond to the input variables of the function considered was added. For the problems of cryptanalysis of generators this method allows to select, on the initial stage of the solving process, the variables corresponding to the secret key as priority variables for decision variable selection procedure. Also some basic constants of the solver were changed. Like most of its analogs Minisat periodically changes the activity of all the variables and clauses in order to increase the priority of selection for variables from the clauses derived in the later steps of the search. Moreover, in 2% of cases the Minisat assigns a value to the variable selected randomly, rather than to the variable with the maximum activity. These heuristics show, on average, good results on a broad set of test examples used in the competitions of SAT solvers. However, for the CNFs encoding problem of cryptanalysis they are, in general, not efficient. In all the experiments described below we use the SAT solver in which periodical lowering of the activity and random selection of variables are prohibited. In total, this simple change led to a substantial increase in efficiency of the SAT solver on cryptographic tests. For example, on the CNFs from the decomposition family constructed in the process of cryptanalysis of the generator A5/1 (see Section 5), the SAT solvers Minisat 1.14.1 and Minisat 2.0 did not cope with the tasks in 10 minutes of work (the computations were interrupted). A modified Minisat-C v1.14.1 solved these problems in less than 0.2 seconds on average (see Table 3).

In the preceding section a general procedure for parallel processing of a list of tasks was described. During this procedure the control process monitors the loading of computing cores and send new tasks to the released cores. In practice, a direct implementation of this scheme leads to an excessive growth of transfer costs, but provides uniform loading of the cores.

The efficiency of a SAT solver in a DCS can be improved by using job batches. Each job batch is a subset of the decomposition family $\Delta_*(C)$. Sending of batches instead of single CNFs allows to reduce the cost of the transfer. We decompose $\Delta_*(C)$ into disjoint sets of job batches. The obtained set of the job batches is considered as a task list where each job batch is a list item. For processing this task list we use the technique described in the previous section.

The fact that a decomposition set is a set of Boolean variables makes the problem of transferring the batches to the cores very simple. Indeed, let $X_* = \{x_{i_1}, \ldots, x_{i_d}\}$ be some decomposition set. And let $M$ be the number of computing cores in the DCS. The core with the number $p \in \{1, \ldots, M\}$ we denote by $e_p$. For the sake of simplicity, assume that $M = 2^k$, $k \in \mathrm{N}_1$, and $k < d$. If we suppose that all the tasks in the decomposition family $\Delta_*(C)$, generated by $X_*$, have approximately equal complexity, then when solving the problem in the DCS each core is going to process approximately the same number of tasks. This means that the decomposition family $\Delta_*(C)$ can be partitioned into $2^k$ subfamilies of equal power and each subfamily can be further processed entirely on the corresponding core. For this purpose select in $X_*$ some subset $X_*^k$ of power $k$ ($X_*^k$ can be formed, for example, by the first $k$ variables from $X_*$). The description of the job batch for a particular $e_p, p \in \{1, \ldots, 2^k\}$, is a binary vector $\alpha_p$ of the length $k$, formed by the values of variables from $X_*^k$. Next, for each $e_p, p = 1, \ldots, 2^k$, we consider the set $\Lambda_p$, consisting from $2^{d-k}$ different vectors of the length $d$ of the form $(\alpha_p|\beta)$, where $\beta$ takes all $2^{d-k}$ possible values from the set $\{0,1\}^{d-k}$.

Each core $e_p, p \in \{1, \ldots, 2^k\}$, receives its job batch from the control process as a vector $\alpha_p$ which is used for constructing the set $\Lambda_p$. A subfamily of the family $\Delta_*(C)$ processed on $e_p$ is obtained as a result of substitutions of the vectors from $\Lambda_p$ to $C$.

### 4.2 Cryptanalysis of some keystream generators on a low-performance computing cluster

It is possible to successfully solve problems of cryptanalysis of some cryptographically weak generators (such as the Geffe generator and the Wolfram generator) on an ordinary personal computer with the use of SAT approach (see [23]).

The use of the results described above enabled to perform a successful parallel logical cryptanalysis of the following generators: the threshold generator (with the key length 80 bits), the summation generator (with the key length 63 bits) and the Gifford generator (with the key length 64 bits). These generators are not considered to be cryptographically strong since there are several known attacks for them. However these attacks are sufficiently different from each other. By contrast, main stages of parallel logical cryptanalysis are the same for all mentioned generators.

For our experiments we used a low-performance computing cluster Blackford (see [24]). The computing node of the Blackford cluster has two processors Intel Xeon Quad-Core E5345 2.33 GHz and 8 GB of RAM. This cluster has 20 nodes of the described configuration.

During the process of cryptanalysis of the threshold and the summation generators it was found out that for the successful solving of these problems it is necessary to include into a decomposition set the variables encoding full initial contents of several LFSRs. The CNF obtained after the substitutions of the values of these variables proved to be very simple for the SAT solver. This fact allows us to calculate the value of the predictive function for the corresponding decomposition quickly. Then, we remove one variable from the decomposition set and calculate the value of the predictive function for the obtained set again by the algorithm described above (Section 3). This procedure is repeated until all variants $\tilde{X}$ from some $\tilde{\Omega}$ are tested. Thus for each generator considered below relationship among all variants of $\tilde{X}$ is subject to the rule

$$\tilde{X}_1 \supset \tilde{X}_2 \supset \ldots \supset \tilde{X}_r = X_* \supset \ldots \supset \tilde{X}_s,$$

$|\tilde{X}_{i+1}| = |\tilde{X}_i| - 1$, $i = 1, \ldots, s-1$ for some $s$.

The threshold generator was proposed by J. Bruer in [25] (see also [26]). This generator contains $R, R \geq 3$ LFSRs, that are shifted simultaneously. On the initial step $\tau = 0$ the bits of the secret key are placed into LFSRs. At each moment of time $\tau, \tau \in \{1, 2, \ldots\}$ output bits of LFSRs are used as arguments of the majority function. The value of the majority function is 1 if the majority of its input bits are 1, and 0 otherwise (see Fig. 2). The output bits of the majority function (at each moment of time) form the keystream.

Parallel logical cryptanalysis was applied to the threshold generator based on five LFSRs given by the following connection polynomials: LFSR 1: $X^{13} + X^{10} + X^8 + X^5 + 1$; LFSR 2: $X^{15} + X^{13} + X^3 + X + 1$; LFSR 3: $X^{16} + X^{13} + X^8 + X^2 + 1$; LFSR 4: $X^{17} + X^6 + X^4 + X^2 + 1$; LFSR 5: $X^{19} + X^{18} + X^{17} + X^{14} + 1$.

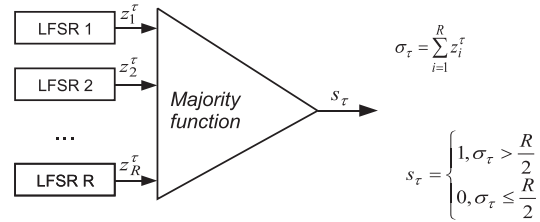

Fig. 2. Threshold generator.

Thus, the length of the secret key in the considered generator was 80 bits. We analyzed first 150 bits of keystream. The initial decomposition set $\tilde{X}_1$ was formed by the variables encoding the contents of the first three LFSRs. The usage of the predictive function technique resulted in the construction of the set $X_*$, which contains the variables encoding full initial contents of the first two LFSRs, plus one variable corresponding to the rightmost bit of the third LFSR. We generated 10 random tests, where "true random" sequences (see [27]) were used as secret keys. The results of cryptanalysis are presented in Table 1 (the correct secret keys were found in all tests). During the cryptanalysis of the threshold generator on a cluster the effect of super-linear acceleration was observed. In our opinion this can be explained by the practical incompleteness of the SAT solvers we use (see Section 3).

The summation generator was proposed by R. Rueppel (see [28], [11]). In this generator, like in the threshold one, each bit of the keystream is an output of a nonlinear function. The function's inputs are outputs of several simultaneously shifted LFSRs (see Fig. 3).

The only difference between this generator and the threshold one is usage of a special summation function (see [11]) instead of the majority function. The summation function uses Carry-register, that is dynamically updated during the work of the generator. The secret key consists of the initial contents of the Carry-register and LFSRs.

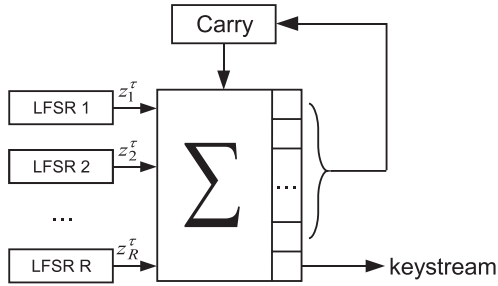

Fig. 3. Summation generator.

On the initial step $\tau = 0$ the bits of the secret key are placed into the Carry-register and LFSRs. At each moment of time $\tau = 1, 2, \ldots$ LFSRs simultaneously output the bits $z_j^\tau, j = 1, \ldots, R$, on the basis of which the following values:

$$S_\tau = \sum_{i=1}^{R} z_i^\tau + C_{\tau-1}, C_\tau = \left\lfloor \frac{S_\tau}{2} \right\rfloor.$$

are calculated. The bit $\varphi_\tau = (S_\tau, mod\, 2)$ is outputted to the keystream, and the binary representation of the number $C_\tau$ is placed into the Carry-register.

Parallel logical cryptanalysis was applied to the summation generator based on the following four LFSRs: LFSR 1: $X^{13}+X^4+X^3+X+1$; LFSR 2: $X^{15}+X^5+X^4+X^2+1$; LFSR 3: $X^{16}+X^6+X^4+X+1$; LFSR 4: $X^{17}+X^6+X^4+X^2+1$.

Thus, two unknown bits of the initial contents of the Carry-register and 61 bits of the initial contents of the LFSRs together form the secret key of the length of 63 bits. We analyzed the first 180 bits of the keystream. Using the technique of predictive functions we constructed a decomposition set $X_*$, formed by the variables encoding the initial contents of the first two LFSRs (28 variables). The results of parallel logical cryptanalysis of this generator are shown in Table 1.

The Gifford generator was developed by a group led by J. Gifford in 1984 (see [29]) and for quite a long time it was used in practice for the transmission of text information. The first successful attack on the Gifford generator is described in the article [30], where a very complicated mathematical apparatus specially developed for cryptanalysis of this generator was presented. A distinctive feature of the Gifford generator is that it does not use LFSRs.

The algorithm of the generator processes the information in groups of 8 bits. On the initial step $\tau = 0$, in the cells $B_1, B_2, \ldots, B_8$ the bytes $b_1^0, \ldots, b_8^0$ of the secret key (the total length is 64 bits) are written (see Fig. 4). The keystream is a sequence of bytes $T_1, T_2, \ldots$, outputted by the generator at the moments of time $\tau \in \{1, 2, \ldots\}$. At each step contents of the cells $B_1, B_2, \ldots, B_8$ is shifted by one byte to the right. At the same time old content of the cell $B_8$ is discarded and the new content of the cell $B_1$ is calculated using the feedback function:

$$b_1^{\tau+1} = f(b_1^\tau, b_2^\tau, b_8^\tau) = b_1^\tau \oplus (>>_1^* (b_2^\tau)) \oplus (<<_1 (b_8^\tau)).$$

Notation $>>_1^*$ means operation of right shift by one bit with the preservation of the high-order bit (sticky right-shift). Notation $<<_1$ means the operation of left shift by one bit with shifting in zero in the low-order bit (zero-fill left shift). I.e. for $B = (x_1, x_2, \ldots, x_8)$ we have

$$>>_1^* (B) = (x_1, x_1, x_2, \ldots, x_7),$$
$$<<_1 (B) = (x_2, x_3, \ldots, x_8, 0).$$

To calculate a byte of a keystream a nonlinear output function $h : \{0,1\}^{32} \to \{0,1\}^8$ is used. This function gets four input bytes (the contents of the cells $B_1, B_3, B_5, B_8$), and outputs a single byte:

$$h(B_1, B_3, B_5, B_8) = ExtractByte_3((B_1|B_3) \times (B_5|B_8)).$$

Here | is a concatenation of the contents of the corresponding cells, and × is an integer multiplication. Thus, the argument of the function $ExtractByte_3$ is a natural number. Binary representation of this number is 32-bit vector. Output of the function $ExtractByte_3(x)$ is the third byte from the left of this vector.

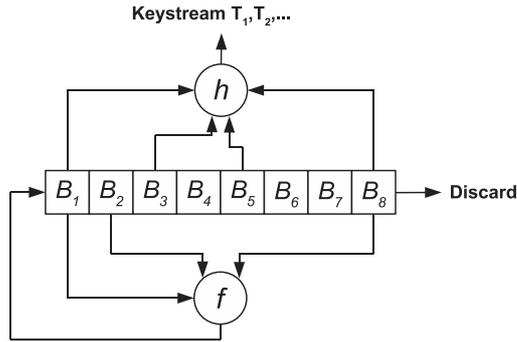

Fig. 4. Gifford generator.

The structure of decomposition sets for Gifford generator is completely different from the structure of these sets for the generators that use LFSRs. Note that the initial content of LFSR defines all of its subsequent states. The structure of decomposition sets in the problems of cryptanalysis of the threshold and summation generators is determined by this very fact. Since the Gifford generator doesn't use LFSRs, the following simple strategy for constructing the initial decomposition set proved to be the most efficient: the set $\tilde{X}_1$ consisted of the variables encoding values of the first 32 bits of the secret key. Then the value of the predictive function was improved according to the method similar to that considered above: each new decomposition set was obtained from a previous one as a result of the removal of some variable. We analyzed first 160 bits of the keystream.

In the end, the technique of predicting functions applied to the Gifford generator gave the following result: if the number of computing cores of a cluster is $2^k$, then one needs to parallelize the SAT problem in $k$ variables corresponding to the first $k$ bits of the secret key. Therefore, in the case of the Gifford generator the power of the decomposition family

TABLE 1
Results of cryptanalysis (10 tests for each generator)

| Generator, secret key length / analyzed fragment of the keystream | Time of parallel solving on 80 cores of the Blackford cluster | | | Time of solving on a single core |
|---|---|---|---|---|
| | Min | Max | Aver | |
| Threshold 5 LFSR, 80 bits / first 150 bits | 10 min. | 46 min. | 27 min. | > 2 days (interrupted) |
| Summation 4 LFSR, 63 bits / first 180 bits | 1 h. 14 min. | 7 h. 43 min. | 3 h. 21 min. | > 2 days (interrupted) |
| Gifford, 64 bits / first 160 bits | 7 min. | 9 h. 53 min. | 4 h. 57 min. | for some tests less than 1 day |

$\Delta_*(C)$ coincides with the number of computing cores. For our experiments we had 80 cores available on the cluster. Since 80 is not a power of 2 the original SAT problem was parallelized in 6 variables which resulted in 64 tasks. To load the remaining cores we divided 15 of the initial 64 SAT problems into two subtasks each, thus obtaining 79 tasks. One remaining core performed the controlling functions.

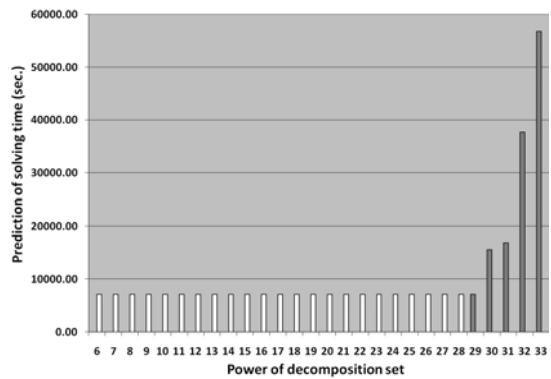

Fig. 5. Fragment of the process of optimization of a predictive function for the threshold generator.

On the Fig. 5 we give an example of a

graph displaying process of optimization of a predictive function for the threshold generator cryptanalysis. The white columns indicate that for the corresponding variants of the decomposition set the calculation of the predictive function was interrupted because of exceeding the current threshold value (see Section 3).

## 5 CRYPTANALYSIS OF THE GENERATOR A5/1 IN A GRID SYSTEM

### 5.1 Construction of a decomposition set

The problem of constructing a good decomposition set for the parallel cryptanalysis of A5/1 proved to be quite nontrivial. We investigated different variants of constructing of decomposition set. First, using some "reasonable" assumptions, we constructed a basic decomposition set $X'$. Then we tried to reduce it applying the predictive function technique.

We propose to include into the decomposition set $X'$ the variables encoding the initial states of the cells of registers, starting with the first cells until the cells containing clocking bits inclusive (corresponding cells in the Fig. 6 are dark shaded). Thus, the decomposition set $X'$ consists of 31 variables:

$$X' = \{x_1, \ldots, x_9, x_{20}, \ldots, x_{30}, x_{42}, \ldots, x_{52}\} \quad (5)$$

This choice is motivated by the following considerations. Assigning values to all variables from $X'$ we determine the exact values of clocking bits for a large number of subsequent states of all three registers. These clocking bits are the most informative because they determine the value of the majority function. The fact that we couldn't further reduce the set $X'$ (5) by applying the predictive function technique was quite unexpected. In our statistical experiments for each variant of decomposition set and the initial fragment of the keystream of some fixed length we constructed random samples of the volume of 1000 CNF. For each such sample we calculated the value of the predictive function (see Section 3). In the Table 3 we show the values of the predictive function calculated for different variants of decomposition sets and keystream lengths. All of the variants of decomposition sets, nevertheless, are conceptually similar to (5). For example, a decomposition set consisting of 30

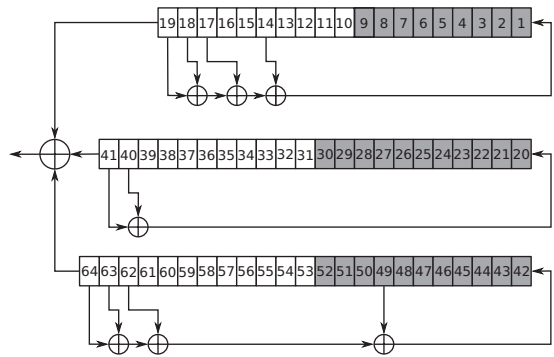

Fig. 6. Scheme of a decomposition set consisting of 31 variables.

variables was formed according to the scheme on Fig. 7.

In all computational experiments SAT problems were solved using a modified variant of the SAT solver Minisat-C v1.14.1 (the details of modification are described in Section 4). As a test platform a single core of the processor Intel E8400 + 2Gb RAM was used.

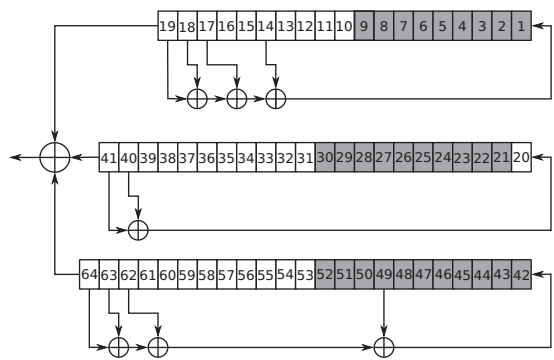

Fig. 7. Scheme of a decomposition set consisting of 30 variables.

Next, we present the prognosis of the time required for solving the problem of cryptanalysis of the generator A5/1 on the computational cluster SKIF MSU "Chebyshev" (see [31]) that consists of 1250 quad-core processors E5472 (the peak performance of the cluster is 60 Tflop/s). Table 2 shows the comparative characteristics of the processor Intel E5472 and the processor Intel E8400.

From this table we can see that the cores of Intel E8400 and Intel E5472 are comparable

13in power (there is only a slight difference in the bus frequency). Because of this we think it is reasonable to use the results of the numerical experiments shown in the Table 3 for the estimation of the time required for the solving the problem of logical cryptanalysis of the generator 5/1 on the cluster SKIF MSU "Chebyshev". In addition, note that the use of the technique of job batches transfer described in the previous section, makes the impact of transfer costs on the overall computation time quite negligible.

TABLE 2
Characteristics of processors

| Processor model | Intel E8400 | Intel E5472 |
|---|---|---|
| Number of cores | 2 | 4 |
| Core frequency | 3.0 GHz | 3.0 GHz |
| Bus frequency | 1333 MHz | 1600 MHz |
| Cache L2 | 6 Mb | 12 Mb |

TABLE 3
Values of the predictive function for a single core of the processor Intel E8400 for the generator A5/1 cryptanalysis (in hundreds millions of seconds)

| Power of the decomposition set | Length of keystream | | | | |
|---|---|---|---|---|---|
| | 128 | 144 | 160 | 176 | 192 |
| 29 | 3.87 | 3.80 | 3.69 | 3.95 | 3.76 |
| 30 | 3.65 | 3.59 | 3.59 | 3.71 | 3.83 |
| 31 | 3.76 | *3.55* | 3.71 | 3.73 | 3.81 |
| 32 | 4.23 | 4.15 | 4.27 | 4.39 | 4.32 |
| 33 | 4.70 | 4.87 | 4.89 | 4.95 | 5.23 |

From all the above, our estimation of the computing time of logical cryptanalysis of A5/1 on the "Chebyshev" cluster is 9–12 hours in average. The corresponding parameters of the decomposition are: the power of decomposition set—31 variables (Fig. 6), the number of analyzed bits of the keystream—144 (first bits of the stream).

### 5.2 The necessity of application of distributed computing technologies to the problem of cryptanalysis of the generator A5/1

The predicted time for solving of this problem on the supercomputer "Chebyshev" shows that even if the cluster is fully dedicated to this task, the process of solving requires considerable time. Exclusive use of public access multiprocessor computing complexes is usually not possible. At the same time, the researchers often have resources of various clusters, Grid systems, high-performance servers at their disposal. The software complex BNB-Grid [32] makes it possible to use such heterogeneous distributed computing resources (called computing nodes) for solving complex computational problems. It has already shown high efficiency in application to several large scale optimization problems [33].

The structure of the computational algorithm for solving the problem of cryptanalysis of the generator A5/1 allows an efficient implementation in parallel and distributed systems. This is possible because subtasks of the decomposition family can be processed independently. Next, we describe the process of solving the problem of cryptanalysis of A5/1 in the BNB-Grid which uses computational resources of several multiprocessor systems.

### 5.3 Computing complex BNB-Grid

Hierarchical structure of BNB-Grid is shown on the Fig. 8. On the top level the object Computing Space Manager (CS-Manager) is located. It decomposes the original problem into subproblems and distributes them among the computing nodes. For each computing node there is a corresponding object of the type Computing Element Manager (CE-Manager). CE-Manager provides communication between CS-Manager and the corresponding computing node and also starts and stops applications on this node. After receiving a task from the CS-Manager, CE-Manager transfers it to the corresponding node and starts MPI application BNB-solver which processes the received task on all available cores.

The core of the system is implemented in Java programming language using the middleware Internet Communication Engine (ICE) [34]—an analog of CORBA. The graphical user interface is also implemented as an ICE-object, indicated on the figure as GUI Manager. ICE-objects are located either on one or on several computers within a local network.

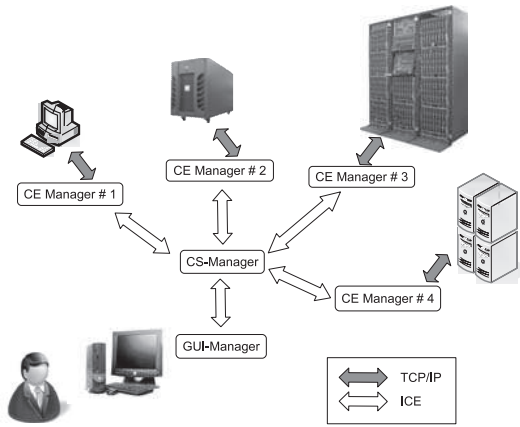

Fig. 8. Organization of computations in BNB-Grid.

Several copies of the BNB-Solver application can be started on one computing node. This approach is often proved to be reasonable for shared access supercomputers working under the control of batch processing systems. In such systems an application requesting a large number of processors can be queued for a long time waiting for an appropriate "window" in the tasks schedule. At the same time, an application requesting a significantly smaller number of processors can be launched earlier.

### 5.4 Parallel solving of SAT problems in BNB-Grid

A module for processing SAT problems on a computing cluster was added to the BNB-Solver. The input data of the control object CS-Manager is a description of the original SAT problem in XML format. CS-Manager decomposes SAT problem for the original CNF $C$ and obtains decomposition family $\Delta_*(C)$. CE-Manager transfer job batches, constructed according to the technique described in the Section 4, to the computing nodes.

For better efficiency of the solving process some reasonable compromise between the number of job batches and the number of tasks in a batch should be achieved. A large number of tasks in a batch allows us to reduce idle time of the processors. However when the number of tasks in a batch is too large its processing time increases greatly. But this is undesirable because an application which requests too many resources of a cluster may be queued for a long time or even will not be run at all.

For solving of the problem of logical cryptanalysis of generator A5/1 in the BNB-Grid a following decomposition was used. The decomposition family itself consisted of $2^{31}$ SAT problems. It was split into $2^{18}$ disjoint subsets (job batches) with $2^{13}$ SAT problems in each of them. These job batches were processed by BNB-solvers.

### 5.5 Computational experiments

In our experiments three test problems of cryptanalysis of generator A5/1 were solved. The computations were carried out on four computing clusters, which characteristics (see Table 4) are taken from the 11-th edition of the list of the most powerful supercomputers in the CIS [35].

The number of simultaneously working computing cores varied in the process of calculations from 0 to 5568, averaging approximately 2–3 thousand cores. For each test computations were stopped after finding the first satisfying assignment. The first test problem was solved (the secret key of the generator was found) in 56 hours, the second and the third—in 25 and 122 hours of Grid system work.

The problem of cryptanalysis of generator A5/1 is also interesting because the same keystream of arbitrary length can be generated from different secret keys. This fact was noted by J. Golic in [36]. We denote these situations as "collisions" using the evident analogy with the corresponding notion from the theory of hash functions. The approach presented in this article allows us to solve the problem of finding all the collisions of the generator A5/1 for a given fragment of a keystream. Using BNB-Grid we found all collisions for one test problem (we analyzed the first 144 bits of keystream). It turned out that there are only three such collisions (see Table 5). Processing this test problem took 16 days of Grid system work.

TABLE 4
Characteristics of computing clusters

| Name | Institution | Processors | Number of cores |
|---|---|---|---|
| MVS-100k | Joint Supercomputer Center of RAS | Xeon E5450 3 GHz | 7920 |
| SKIF-MSU Chebyshev | Moscow State University | Xeon E5472 3 GHz | 5000 |
| Cluster of RRC | RRC Kurchatov Institute | Xeon 5345 2.33 GHz | 3456 |
| BlueGene P | Moscow State University | Power PC 850 MHz | 8192 |

TABLE 5
Original key and collisions of generator A5/1
(in hexadecimal format)

| | LFSR 1 $x_1,\ldots,x_{19}$ | LFSR 2 $x_{20},\ldots,x_{41}$ | LFSR 3 $x_{42},\ldots,x_{64}$ |
|---|---|---|---|
| **orig. key** | 2C1A7 | 3D35B9 | EEAF2 |
| **collision** | 2C1A7 | 3E9ADC | EEAF2 |
| **collision** | 2C1A7 | 3D35B9 | 77579 |

## 6 CONCLUSION

In this work a parallel technology for solving inversion problems for discrete functions computable in polynomial time is presented. This technology is based on a reduction of the considered problems to SAT problems. Using the information about the input variables of the considered function we construct a decomposition of the corresponding SAT problem into a family of subproblems. Then, this family is processed as a parallel task list in a distributed computing environment. To construct a good decomposition we use technique of optimization of a special predictive function.

The technology presented in the work was tested on problems of cryptanalysis of some keystream generators (threshold, summation, Gifford generator). These problems were solved on a low-performance computing cluster. For solving the problem of cryptanalysis of the keystream generator A5/1 a Grid-system was specially constructed. Although this task was quite time-consuming (from 1 to 16 days), all the tests were correctly solved. Thus, the possibility of cryptanalysis of A5/1 in public access computing environments without using of special computing architectures (like for example in [37]) was experimentally confirmed.

An implementation of the described technology in more powerful Grid-systems or on high performance clusters makes the problem of cryptanalysis of A5/1 solvable within several hours. We particularly emphasize that our approach makes it possible to find a secret key using only very small fragment of keystream (first 144 bits). This result can be considered as one of many votes against the widespread use of the cipher A5/1. On the webpage [38] CNFs encoding the problem of cryptanalysis of the generator A5/1 are available.

Let us emphasize that the main purpose of the present work was the development of a technology for solving inversion problems for polynomially computable discrete functions in distributed computing environments. The results of cryptanalysis presented in the article were not the ultimate goal itself and should be considered only as arguments for the efficiency of the proposed technology.

### ACKNOWLEDGMENTS

The authors would like to thank Alexei Hmelnov, Stepan Kochemazov and Alexey Ignatiev (ISDCT SB RAS) for help and numerous valuable advices.